%% file: DPF_proceeding.tex
\documentclass[12pt]{article}
\usepackage{graphicx}

\usepackage{ifthen} 
\newboolean{pdflatex}
\setboolean{pdflatex}{true} 

\newboolean{articletitles}
\setboolean{articletitles}{true} 

\newboolean{uprightparticles}
\setboolean{uprightparticles}{false} 

\newboolean{inbibliography}
\setboolean{inbibliography}{false}
\input{preamble}
\usepackage{longtable}

\newcommand\pubnumber{DPF2015-24}
\newcommand\pubdate{\today}

\def\ferrara{\textit{Sezione INFN di Ferrara, Ferrara, Italy,\\ European Organization for Nuclear Research (CERN), Geneva, Switzerland, Universit\`a di Ferrara, Ferrara, Italy}}
\def\support{\footnote{On behalf of the LHCb Collaboration.}}

\def\Title#1{\begin{center} {\Large #1 } \end{center}}
\def\Author#1{\begin{center}{ \sc #1} \end{center}}
\def\Address#1{\begin{center}{ \it #1} \end{center}}

\newcommand\pubblock{\rightline{\begin{tabular}{l} \pubnumber\\
         \pubdate  \end{tabular}}}
\newenvironment{Abstract}{\begin{quotation}  }{\end{quotation}}
\newenvironment{Presented}{\begin{quotation} \begin{center} 
             PRESENTED AT\end{center}\bigskip 
      \begin{center}\begin{large}}{\end{large}\end{center} \end{quotation}}

\begin{document}

\begin{titlepage}
\pubblock

\vfill
\Title{New results on semileptonic b decays from LHCb}
\vfill
\Author{ Marco Fiore\support}
\Address{\ferrara}
\vfill
\begin{Abstract}
We report new measurements, based on the Run I dataset collected by the LHCb experiment, of the \Bz mixing frequency \dmd and the CKM matrix element $|V_{ub}|$ using semileptonic b decays. The former is the most precise measurement ever performed; the latter represents the first determination of $|V_{ub}|$ using a baryonic decay, and adds an important constraint on a possible explanation for the discrepancy between exclusive and inclusive determinations of this observable.
\end{Abstract}
\vfill
\begin{Presented}
DPF 2015\\
The Meeting of the American Physical Society\\
Division of Particles and Fields\\
Ann Arbor, Michigan, August 4--8, 2015\\
\end{Presented}
\vfill
\end{titlepage}
\def\thefootnote{\fnsymbol{footnote}}
\setcounter{footnote}{0}

\section{Introduction}
The LHCb detector~\cite{Detector} was designed and built to study decays of hadrons containing $b$ and $c$ quarks, to search for indirect evidence of new physics beyond the Standard Model. Due to their relatively high branching fractions, semileptonic decays of B hadrons can be exploited to perform precise measurements of observables in the B-physics sector. \\
During 2011 and 2012, in the so called Run I of the LHC, the LHCb experiment collected an integrated luminosity of 3 fb$^{-1}$ at center of mass energies of $\sqrt{s} = 7$ and $\sqrt{s} = 8 \tev$.
Two measurements performed using the Run I dataset are reported here: the measurement of the \Bz meson mixing frequency \dmd using $\Bz \rightarrow D^{(*)-} \mu+ \nu_{\mu}$ decays~\cite{Dmd} and the measurement of the CKM matrix element $|V_{ub}|$ using $\Lambda_b \rightarrow p \mu \nu_{\mu}$ decays~\cite{Vub}. \\

\section{\boldmath{\dmd} with $\Bz \rightarrow D^{(*)-} \mu^+ \nu_{\mu}$ decays}
The time evolution of a system consisting of two neutral mesons is governed by the Schr{\"o}dinger equation.
In this particular case, the \Bz system is chosen, but the equations are equally valid for the other neutral meson systems.
Inserting a term to describe the decay of the two mesons, the Schr{\"o}dinger equation is written as
\begin{equation}
  \label{eq:Schrodinger}
  i \frac{\rm d}{{\rm d}t} \left( \begin{array}{c}
    \ket{B^0(t)}  \\
    \ket{\overline{B}^0(t)} \end{array} \right) =
  \left(M-\frac{i}{2}\Gamma\right) \left( \begin{array}{c}
    \ket{B^0(t)}  \\
    \ket{\overline{B}^0(t)}  \end{array} \right) \;\;\; ,
\end{equation}
where $M$ and $\Gamma$ are hermitian matrices, which describe the mass
(dispersion) and decay (absorption) of the B system, respectively. The eigenstates of the
Schr{\"o}dinger equation Eq.~(\ref{eq:Schrodinger}) are the mass eigenstates of
neutral B mesons, defined as
\begin{equation}
  \ket{B_{H,L}} = p \ket{B^0} \mp q \ket{\overline{B}^0} \ ,
\end{equation}
where $B_H$ is the {\em heavy} eigenstate and $B_L$ is the {\em light}
eigenstate. The difference in mass and decay rate are defined as
\begin{align}
  \Delta m = m_H - m_L , \nonumber \\
  \Delta \Gamma = \Gamma_L - \Gamma_H \ . 
\end{align}
In the absence of CP violation in mixing, the heavy mass eigenstate is CP-odd and the light one CP-even. In semileptonic $\Bz \rightarrow D^{(*)-} \mu^+ \nu_{\mu}$ decays, the final state determines the flavour of the B meson at the time of decay.
Furthermore, when the \D meson decays in a Cabibbo-favoured mode, it can be safely assumed that there is no direct CP violation.
In such flavour-specific decays, the time dependent decay rates are
\begin{align}
  N(\Bz \rightarrow D^{(*)-} \mu^+ \nu_{\mu})(t) \propto e^{-\Gamma t} [\cosh(\Delta \Gamma t/2) +
\cos (\Delta m t)] \ ,\nonumber \\
  N(\Bz \rightarrow \overline{B}^0 \rightarrow D^{(*)+} \mu^- \overline{\nu_{\mu}})(t) \propto  e^{-\Gamma t} [ \cosh(\Delta \Gamma t/2) - \cos (\Delta m t) ] \ .
\end{align}
Neglecting CP violation in mixing and assuming $ \Delta \Gamma = 0$ in the ${B}^0$ system, the flavour asymmetry between unmixed ($N^{\rm{unmix}}$) and mixed ($N^{\rm{mix}}$) events is
\begin{equation}\label{eq:asym}
A(t) = \frac{N^{\rm{unmix}}(t) - N^{\rm{mix}} (t)}{N^{\rm{unmix}}(t) + N^{\rm{mix}}(t)} = \cos (\Delta m t)  \ .
\end{equation}
From this asymmetry it is possible to extract the value of \dmd via a time-dependent fit.\\

\begin{figure}[]
\begin{center}
\includegraphics[width=0.9\linewidth]{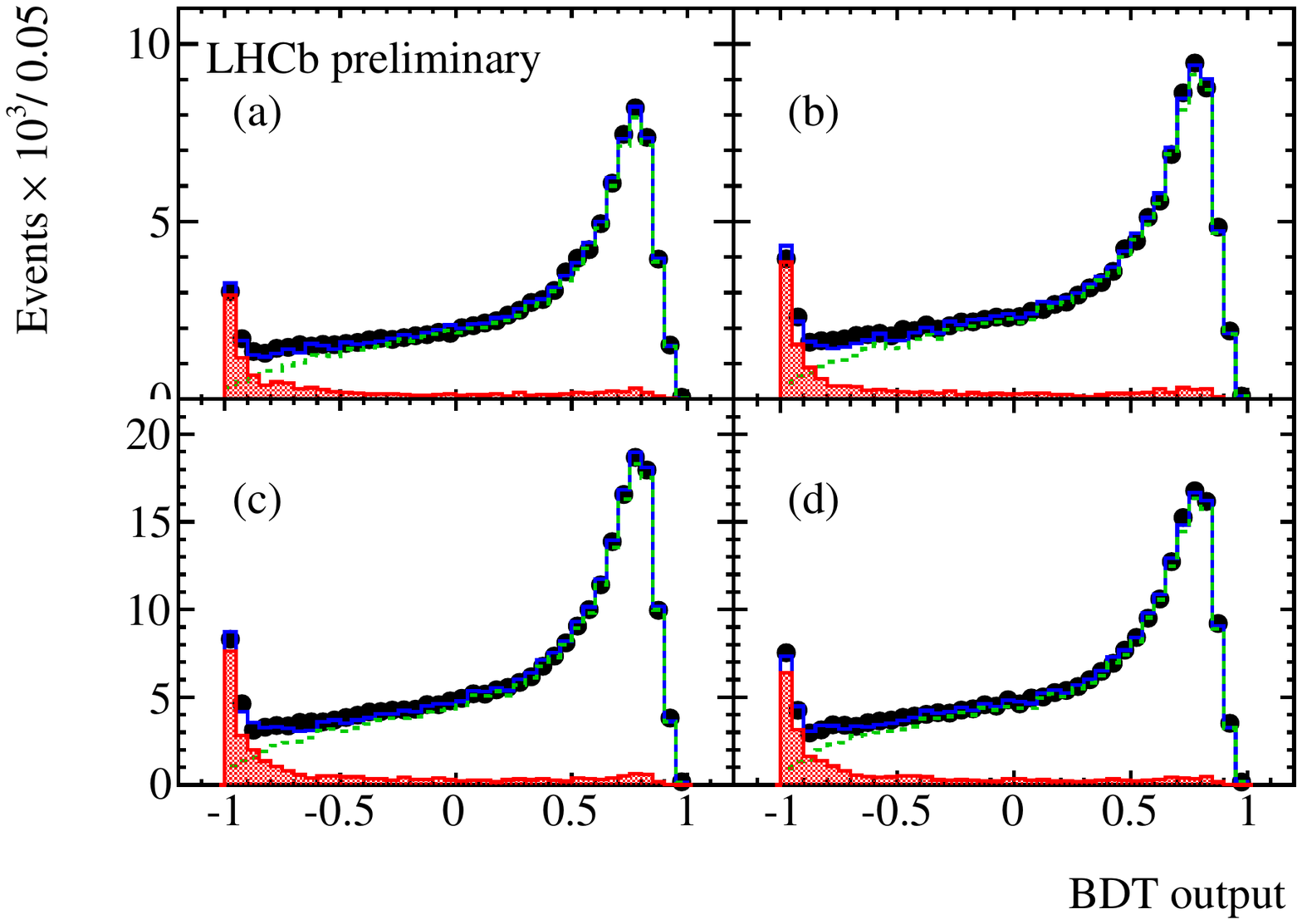}
\end{center}
\caption{Fit to the BDT output distribution for $\Bz \rightarrow D^{*-} \mu^+ \nu_{\mu}$ $\sqrt{s} = 8$ TeV data in the four tagging categories.}
\label{fig:BDT}\end{figure}

\begin{figure}[]
\begin{center}
\includegraphics[width=0.49\linewidth]{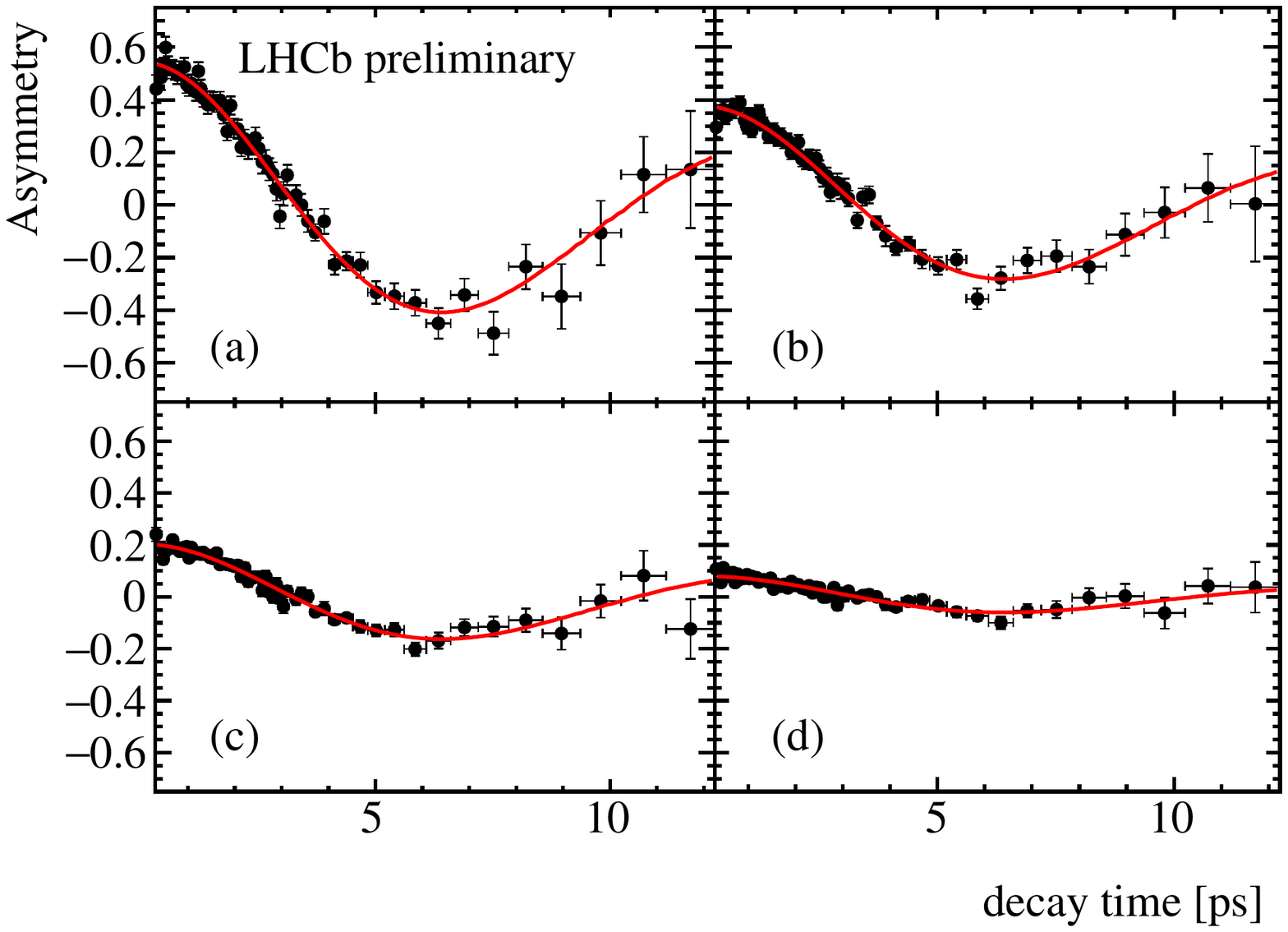}\put(-20,145){(a)}
\includegraphics[width=0.49\linewidth]{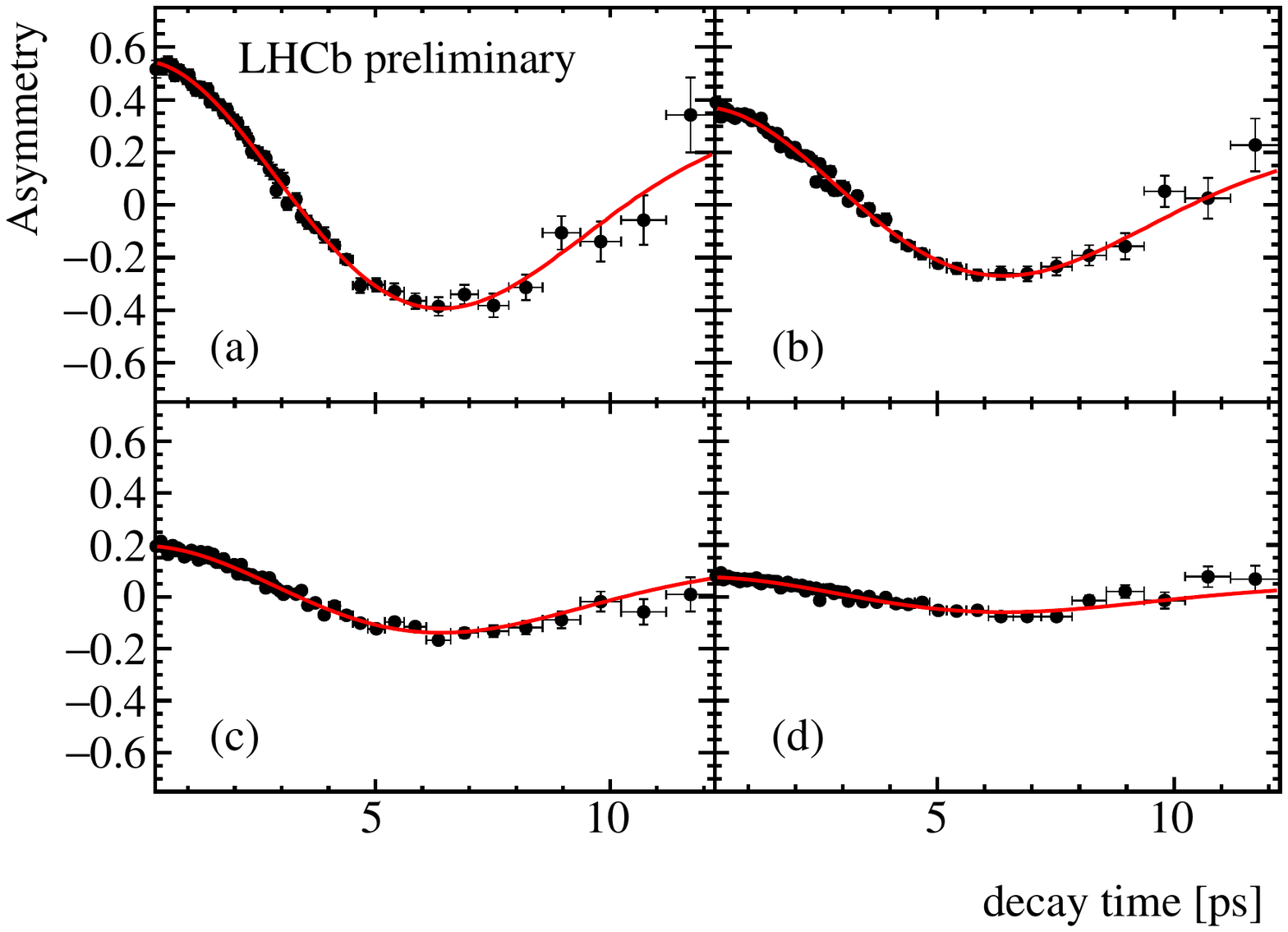}\put(-20,145){(b)}
 \vspace*{0.5cm}
\includegraphics[width=0.49\linewidth]{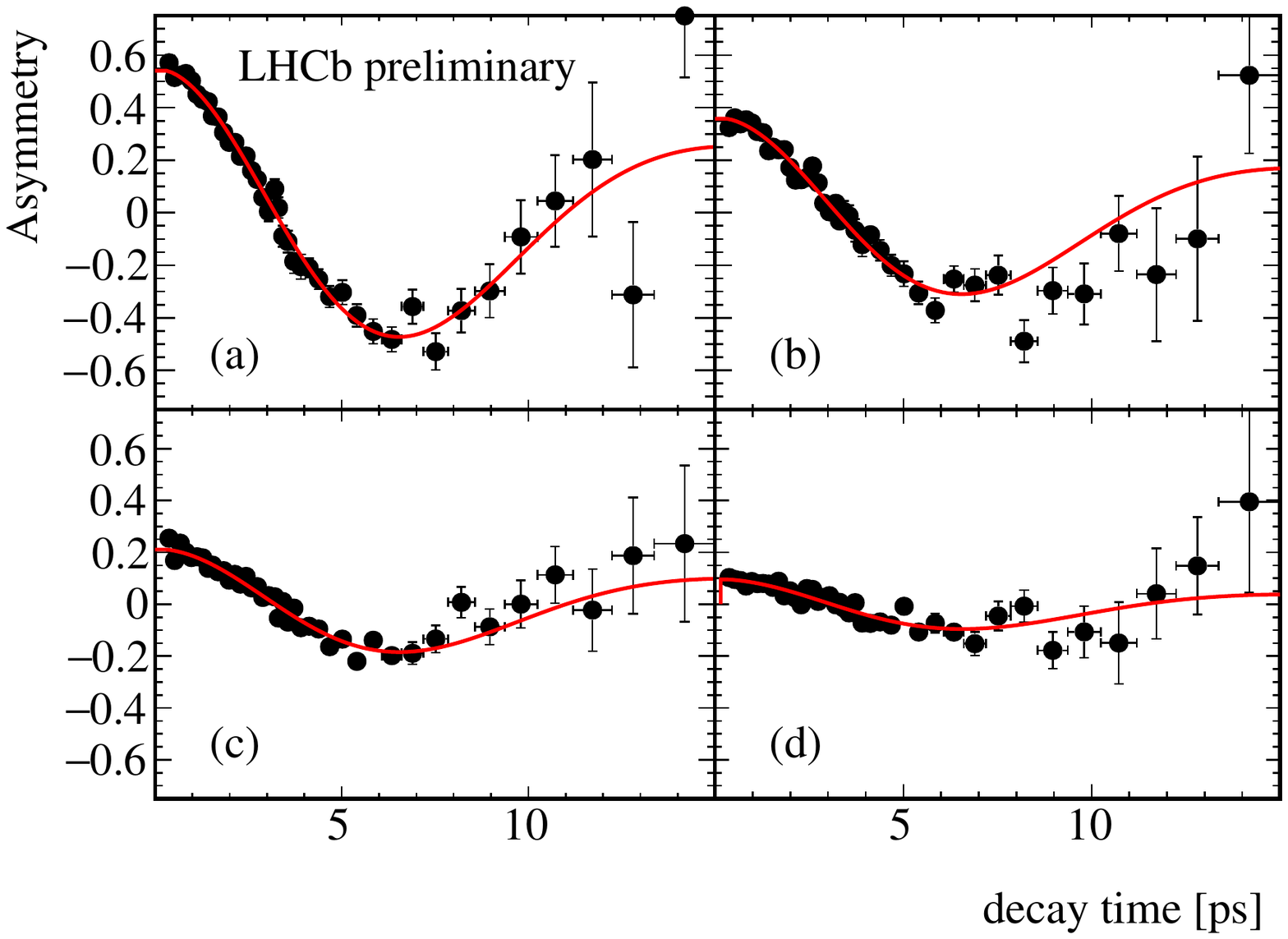}\put(-20,145){(c)}
\includegraphics[width=0.49\linewidth]{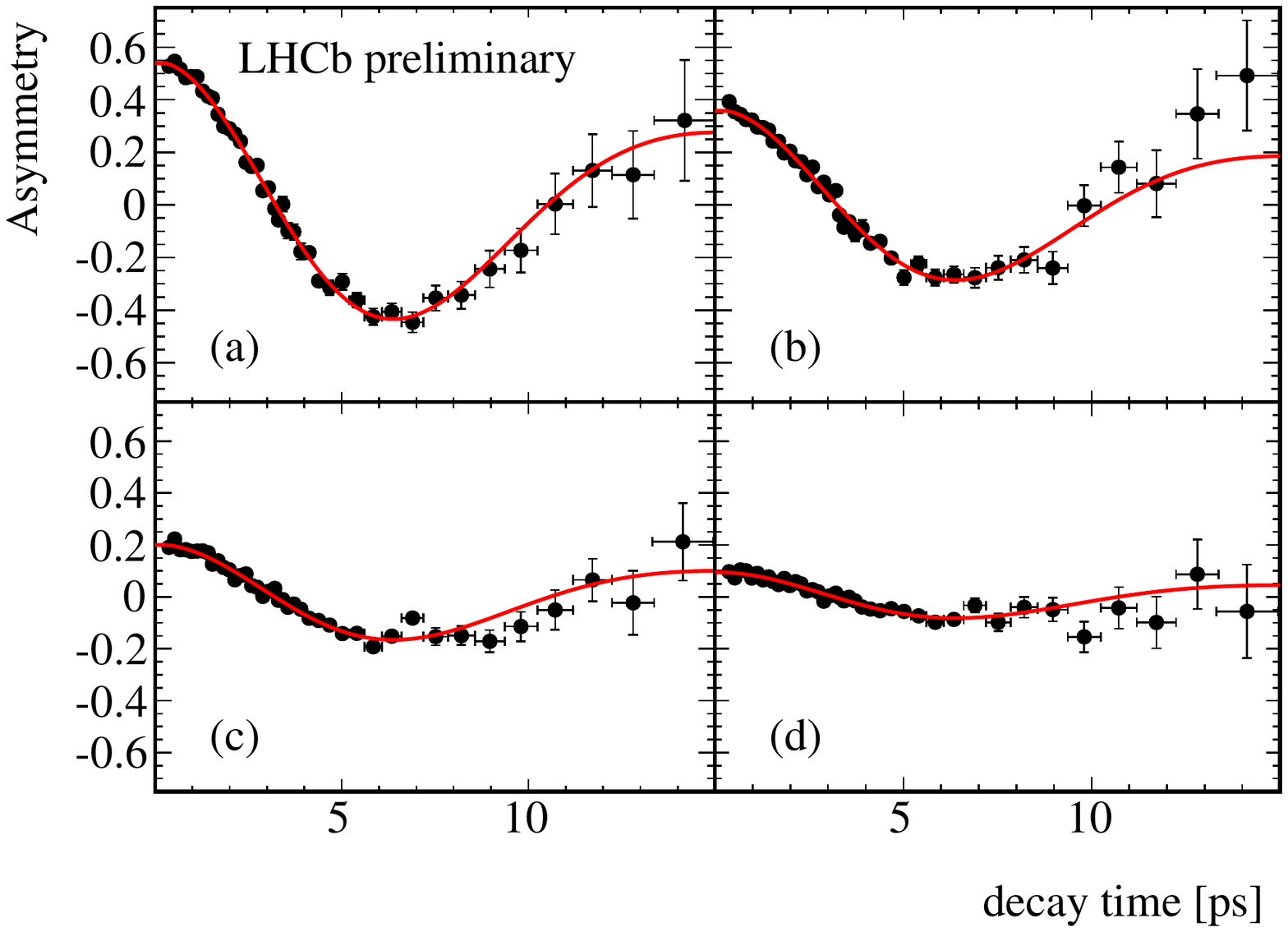}\put(-20,145){(d)}
\end{center}
\caption{Mixing asymmetries projections for $\Bz \rightarrow D^{-} \mu^+ \nu_{\mu}$ and $\Bz \rightarrow D^{*-} \mu^+ \nu_{\mu}$ with $\sqrt{s} = 7$ TeV ((a) and (c)) and $\sqrt{s} = 8$ TeV ((b) and (d)) data respectively.}
\label{fig:asym_res}\end{figure}

\begin{figure}[]
\begin{center}
\includegraphics[width=0.75\linewidth]{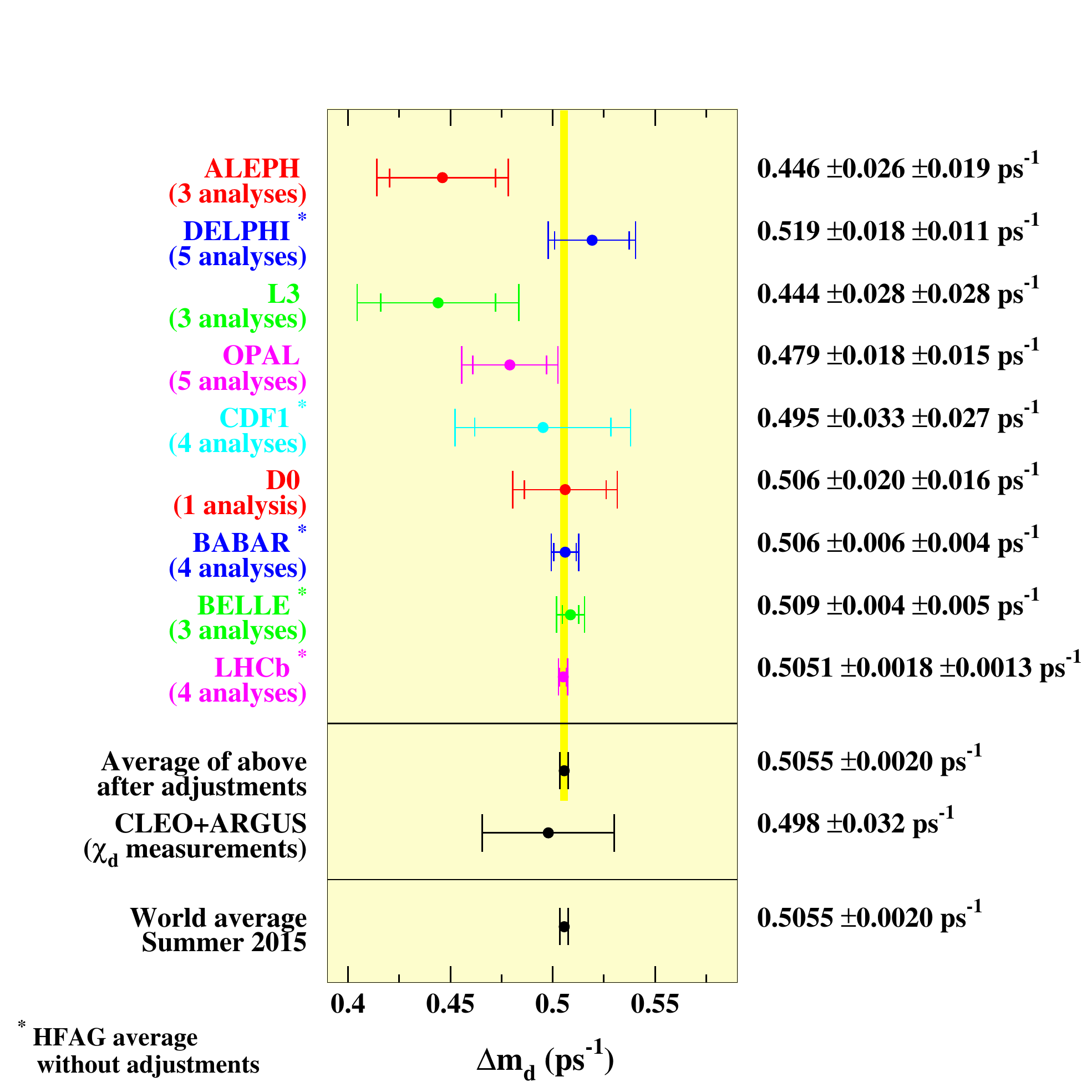}
\end{center}
\caption{World Average on \dmd with the inclusion of the measurement reported here. Previous measurements from other experiments are shown as well.}
\label{fig:NWA}\end{figure}

The flavour at production can be determined using flavour tagging methods~\cite{Tagging}. Different algorithms (taggers) are used to provide the tagging decision and the mistag probability, the latter is floated in the fit and is also used to split the sample in four categories of increasing mistag to increase the statistical precision. \\
To compute the asymmetry in equation~\ref{eq:asym} it is needed to calculate the decay time of the \Bz, defined as
\begin{equation}
  \label{eq:decaytime}
   t = t_{\rm{meas}} \cdot k = \frac{M_{\Bz} \cdot L}{p_{D^{(*)}\mu}\cdot c} \cdot k \ ,
\end{equation}
where $M_{\Bz}$ is the mass of the \Bz, $L$ its decay length, $p_{D^{(*)}\mu}$ the reconstructed momentum and $c$ the speed of light. The parameter $k$, defined as the ratio between the reconstructed and the true \Bz momentum:
\begin{equation}
  \label{eq:kfactor}
   k = \frac{p_{D^{(*)}\mu}}{p^{\rm true}_{\Bz}} \ ,
\end{equation}
is obtained from Monte Carlo, and is used to correct the measured decay time $t_{\rm{meas}}$ and describe its resolution. This correction is needed because of the missing neutrino, which makes impossible to measure precisely the momentum of the \Bz meson. The k-factor is parameterised with a fourth-order polynomial as a function of the $D^{(*)}\mu$ invariant mass.\\
The dominant background for this analysis is due to $B^+ \rightarrow D^{(*)-} \mu^+ \pi^+ \nu_{\mu} X $ decays. It is reduced through a Boosted Decision Tree (BDT) that exploits topological differences between signal and background. The BDT is able to retain 90\% of the signal and reduce the $B^+$ background by 70\%. The fraction of the remaining $B^+$ background is determined from fits to the BDT output distribution. An example of this fit, performed on $\Bz \rightarrow D^{*-} \mu^+ \nu_{\mu}$ $\sqrt{s} = 8$ \tev data, is shown in Figure~\ref{fig:BDT}.
\\
The value of \dmd is determined by simultaneous binned maximum-likelihood fits to the time distributions for mixed and unmixed events, in four mistag categories, separately for the two decay channels and $\sqrt{s} = 7$ and $\sqrt{s} = 8$ \tev datasets. The resulting asymmetries are reported in Figure~\ref{fig:asym_res}.
\\
The value of \dmd is obtained by combining the results on the two channels: 
\begin{align}
\Delta m_d = (503.6 \pm 2.0 (\rm{stat}) \pm 1.3 (\rm{syst})) \rm{ns}^{-1} .
\end{align}
This is the most precise measurement of \dmd to date. \\
Several sources of systematic uncertainties, due e.g. to the k-factor method, the $B^+$ and other backgrounds' fractions and decay time-dependent selection efficiencies, are studied thoroughly with parameterised simulation. \\
Figure~\ref{fig:NWA} gives the status of \dmd determinations from different experiments, and the resulting average, indicated by the vertical yellow band. \\ 
The inclusion of this measurement changes the World Average~\cite{HFAG} from $\dmd = (510 \pm 3)\rm{ns}^{-1}$ to $\dmd = (505.5 \pm 2.0)\rm{ns}^{-1}$, giving an improvement of 33\% in precision.\\
\section{Determination of \boldmath{$|V_{ub}|$} with $\Lambda^0_b \rightarrow p \mu \nu$ decays}
Among the Cabibbo-Kobayashi-Maskawa (CKM)~\cite{Cabibbo}~\cite{KM} matrix elements, $|V_{ub}|$ is the least precisely measured. This, in addition to a discrepancy at the 3$\sigma$ level between determinations obtained with inclusive and exclusive charmless semileptonic decays, motivates an independent $|V_{ub}|$ measurement with data from the LHCb detector. Several explanations have been proposed to explain this discrepancy, such as the presence of a right-handed (vector plus axial-vector) coupling as an extension of the SM beyond the left-handed (vector minus axial-vector) $W$ coupling~\cite{RH}.\\
The magnitude of $|V_{ub}|$ can be determined from 
\begin{align}
\frac{|V_{ub}|^2}{|V_{cb}|^2} =
\frac{\mathcal{B}(\Lambda^0_{b} \rightarrow p \mu \nu)_{q^2 > 15~\rm{GeV}^2}}{\mathcal{B}(\Lambda^0_{b} \rightarrow \Lambda_{c} (\rightarrow p K \pi) \mu \nu)_{q^2 > 7~\rm{GeV}^2}} R_{FF}
\end{align}
where $\mathcal{B}$ denotes the branching fractions measured in kinematic regions defined by a cut in the squared di-lepton invariant mass $q^2$, and $R_{FF}$ is a ratio of the relevant form factors, calculated using Lattice QCD (LQCD)~\cite{LQCD}. The value $|V_{cb}| = (39.5 \pm 0.8) \times 10^{-3}$, obtained from exclusive decays~\cite{PDG}, is used to obtain $|V_{ub}|$. The normalisation to the $\Lambda^0_{b} \rightarrow \Lambda_{c} \mu \nu$ decay cancels out many experimental uncertainties, including the one from the total production rate of $\Lambda^0_{b}$ baryons.\\ 
\begin{figure}[h!]
\begin{center}
\includegraphics[width=0.49\linewidth]{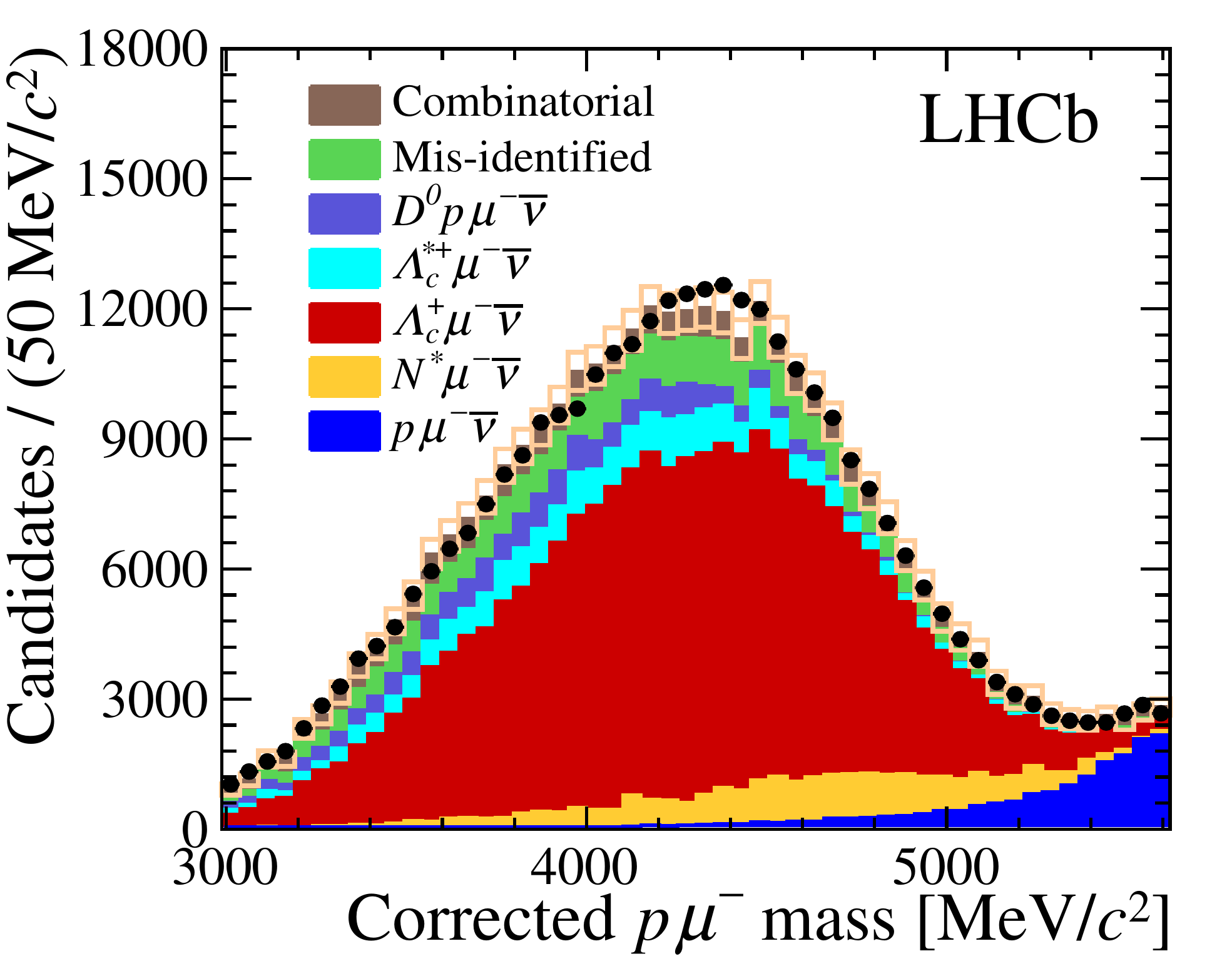}\put(-10,172){(a)}
\includegraphics[width=0.49\linewidth]{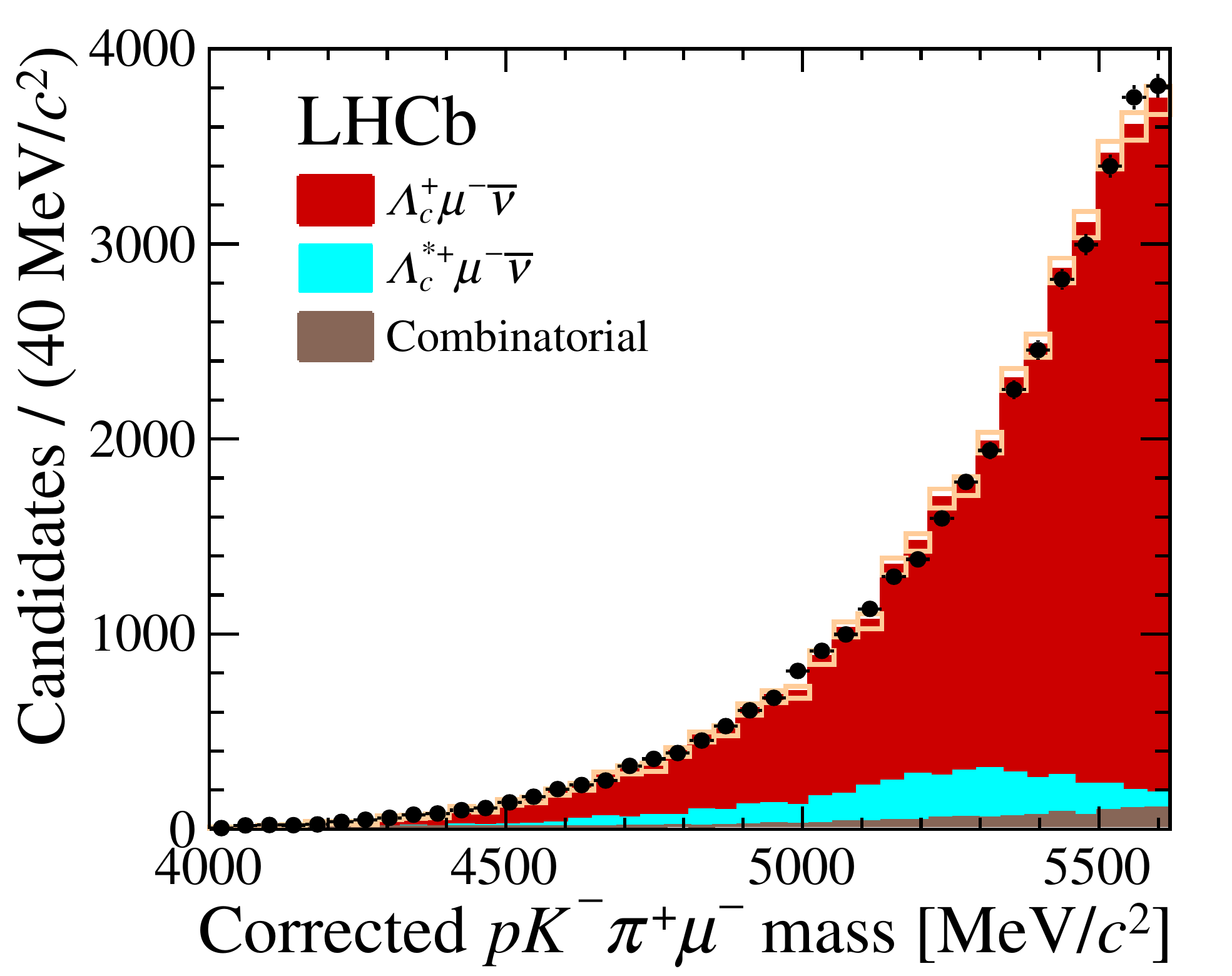}\put(-10,172){(b)}
\end{center}\label{fig:Yields}
\caption{Corrected mass fits for determining signal yields, for signal (a) and normalisation (b).}
\end{figure}
\begin{figure}[h!]
\begin{center}
\includegraphics[width=0.7\linewidth]{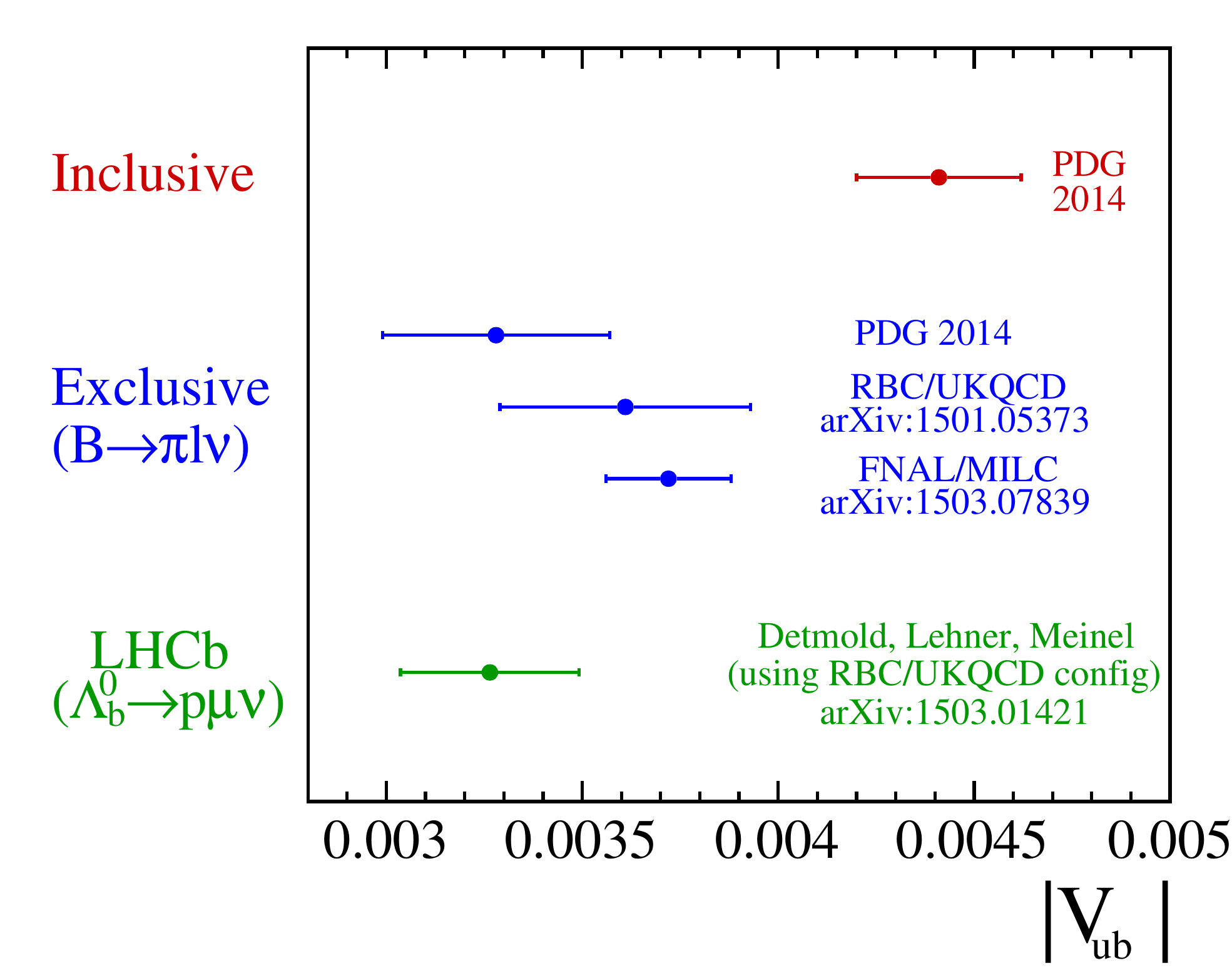}
\end{center}\label{fig:InclExcl}
\caption{Comparison of this measurement (green) with the current averages for exclusive (blue) and inclusive (red) $|V_{ub}|$ results made by the PDG.}
\end{figure}
\begin{figure}[h!]
\begin{center}
\includegraphics[width=0.7\linewidth]{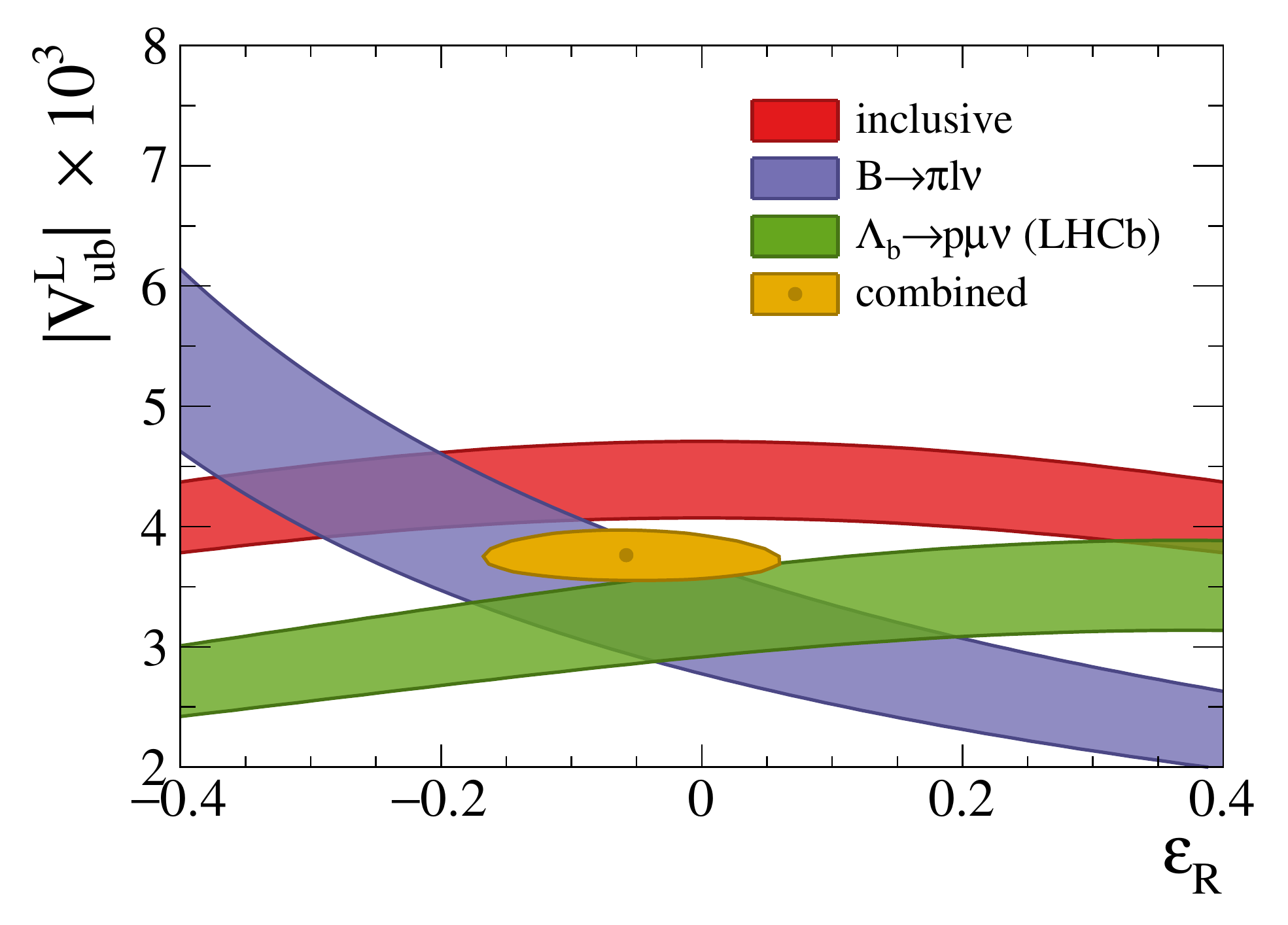}
\end{center}\label{fig:RH}
\caption{Experimental constraints on the left-handed coupling $|V^L_{ub}|$ and the fractional right-handed coupling $\epsilon_R$.}
\end{figure}
The selection for this analysis makes use of a Boosted Decision Tree (BDT) to isolate background with additional tracks coming from the same vertex as the signal candidate. The $\Lambda^0_{b}$ mass is reconstructed using the corrected mass, defined as
\begin{align}
M_{corr} = \sqrt{M^2_{X\mu} + p^2_T} + p_T  ,
\end{align}
where $M^2_{X\mu}$ is the visible mass and $p_T$ is the missing momentum transverse to the $\Lambda^0_{b}$ direction of flight. Candidates with an uncertainty of less than 100 \mevcc on the corrected mass are selected for the $\Lambda^0_{b} \rightarrow p \mu \nu$ decay. This improves significantly the separation between signal and background for the selected candidates.
The signal ($\Lambda^0_{b} \rightarrow p \mu \nu$) and normalisation ($\Lambda^0_{b} \rightarrow \Lambda_{c} \mu \nu$) yields are extracted from a fit to the corrected mass.
The fits, shown in Figure~4, give the following yields: 
\begin{align}
N_{\Lambda^0_b \rightarrow p \mu} = 17687 \pm 733 ,\nonumber \\
\hspace{1cm}
N_{\Lambda_c\rightarrow p K \pi \mu } = 34255 \pm 571.
\end{align}
This is the first observation of the $\Lambda^0_{b} \rightarrow p \mu \nu$ decay. \\
Taking the exclusive determination of $|V_{cb}|$ from the PDG~\cite{PDG}, the magnitude of $V_{ub}$ is
\begin{align}
|V_{ub}| = (3.27 \pm 0.23) \times 10^{-3} , 
\end{align}
which is in agreement with the exclusively measured world average but disagrees with the inclusive measurement, as can be seen in Figure~5.
As mentioned before, the determination of $|V_{ub}|$ from the measured ratio of branching fractions depends on the size of a possible right-handed coupling. This can be clearly seen in Figure~6, which shows the experimental constraints on the left-handed coupling $|V^L_{ub}|$ and the fractional right-handed coupling added to the SM $\epsilon_R$ for different measurements. The overlap of the bands from previous measurements suggested a significant right-handed coupling but the inclusion of this measurement does not support that hypothesis.

\section{Conclusions}
Using the Run I dataset, LHCb performed important measurements in the semileptonic b decays sector.
The measurement of the \Bz-\Bzb mixing frequency \dmd is the most precise performed so far, improving the world average precision by one third.
The $|V_{ub}|$ measurement is the first to be performed using a baryonic decay ($\Lambda^0_{b} \rightarrow p \mu \nu$), which is observed for the first time. The deviation between exclusive and inclusive measurements remains and will require additional experimental and theoretical studies.



\end{document}

%% file: preamble.tex
\usepackage{microtype}
\usepackage{lineno}  
\usepackage{xspace} 
\usepackage{caption} 

\usepackage{graphicx}  
\usepackage{color}
\usepackage{colortbl}
\graphicspath{{./figs/}} 

\usepackage{amsmath} 
\usepackage{amssymb}
\usepackage{amsfonts}
\usepackage{upgreek} 

\newcommand*\patchAmsMathEnvironmentForLineno[1]{%
\expandafter\let\csname old#1\expandafter\endcsname\csname #1\endcsname
\expandafter\let\csname oldend#1\expandafter\endcsname\csname
end#1\endcsname
 \renewenvironment{#1}%
   {\linenomath\csname old#1\endcsname}%
   {\csname oldend#1\endcsname\endlinenomath}%
}
\newcommand*\patchBothAmsMathEnvironmentsForLineno[1]{%
  \patchAmsMathEnvironmentForLineno{#1}%
  \patchAmsMathEnvironmentForLineno{#1*}%
}
\AtBeginDocument{%
\patchBothAmsMathEnvironmentsForLineno{equation}%
\patchBothAmsMathEnvironmentsForLineno{align}%
\patchBothAmsMathEnvironmentsForLineno{flalign}%
\patchBothAmsMathEnvironmentsForLineno{alignat}%
\patchBothAmsMathEnvironmentsForLineno{gather}%
\patchBothAmsMathEnvironmentsForLineno{multline}%
\patchBothAmsMathEnvironmentsForLineno{eqnarray}%
}

\usepackage{hyperref}    
\usepackage[all]{hypcap} 

\input{lhcb-symbols-def} 

\usepackage{cite} 
\usepackage{mciteplus}

%% file: lhcb-symbols-def.tex
\def\MagUp {\mbox{\em Mag\kern -0.05em Up}\xspace}

\ifthenelse{\boolean{uprightparticles}}%
{

 \def\PDelta      {\ensuremath{\Delta}\xspace}                 
 \def\PXi      {\ensuremath{\Xi}\xspace}                 
 \def\PLambda      {\ensuremath{\Lambda}\xspace}                 
 \def\PSigma      {\ensuremath{\Sigma}\xspace}                 
 \def\POmega      {\ensuremath{\Omega}\xspace}                 
 \def\PUpsilon      {\ensuremath{\Upsilon}\xspace}                 
 
 \def\PB      {\ensuremath{\mathrm{B}}\xspace}                 
                  
 \def\PD      {\ensuremath{\mathrm{D}}\xspace}

 \def\PK      {\ensuremath{\mathrm{K}}\xspace}

 \def\Pd      {\ensuremath{\mathrm{d}}\xspace}

 \def\Pi      {\ensuremath{\mathrm{i}}\xspace}

}
{

 \mathchardef\PDelta="7101
 \mathchardef\PXi="7104
 \mathchardef\PLambda="7103
 \mathchardef\PSigma="7106
 \mathchardef\POmega="710A
 \mathchardef\PUpsilon="7107
                  
 \def\PB      {\ensuremath{B}\xspace}                 
                  
 \def\PD      {\ensuremath{D}\xspace}

 \def\PK      {\ensuremath{K}\xspace}

 \def\Pd      {\ensuremath{d}\xspace}

 \def\Pi      {\ensuremath{i}\xspace}

}

\makeatletter
\ifcase \@ptsize \relax 
\newcommand{\miniscule}{\@setfontsize\miniscule{4}{5}}}
\or 
\newcommand{\miniscule}{\@setfontsize\miniscule{5}{6}}}
\or 
\newcommand{\minescule}{\@setfontsize\minescule{6}{7}}
\fi
\makeatother

\DeclareRobustCommand{\optbar}[1]{\shortstack{{\miniscule (\rule[.5ex]{1.25em}{.18mm})}
  \\ [-.7ex] $#1$}}

\def\dquark    {{\ensuremath{\Pd}}\xspace}

  \def\Kbar    {{\kern 0.2em\overline{\kern -0.2em \PK}{}}\xspace}

\def\KorKbar    {\kern 0.18em\optbar{\kern -0.18em K}{}\xspace}

  \def\Dbar    {{\kern 0.2em\overline{\kern -0.2em \PD}{}}\xspace}
\def\D       {{\ensuremath{\PD}}\xspace}

\def\DorDbar    {\kern 0.18em\optbar{\kern -0.18em D}{}\xspace}

\def\B       {{\ensuremath{\PB}}\xspace}
\def\Bbar    {{\ensuremath{\kern 0.18em\overline{\kern -0.18em \PB}{}}}\xspace}

\def\BorBbar    {\kern 0.18em\optbar{\kern -0.18em B}{}\xspace}
\def\Bz      {{\ensuremath{\B^0}}\xspace}
\def\Bzb     {{\ensuremath{\Bbar{}^0}}\xspace}

  \def\Y#1S{\ensuremath{\PUpsilon{(#1S)}}\xspace}

\def\Lbar        {{\ensuremath{\kern 0.1em\overline{\kern -0.1em\PLambda}}}\xspace}
\def\LorLbar    {\kern 0.18em\optbar{\kern -0.18em \PLambda}{}\xspace}

\newcommand{\dmd}{{\ensuremath{\Delta m_{\dquark}}}\xspace}

\def\AT#1     {\ensuremath{A_{\mathrm{T}}^{#1}}\xspace}           

\def\C#1      {\ensuremath{\mathcal{C}_{#1}}\xspace}                       
\def\Cp#1     {\ensuremath{\mathcal{C}_{#1}^{'}}\xspace}                    
\def\Ceff#1   {\ensuremath{\mathcal{C}_{#1}^{\mathrm{(eff)}}}\xspace}        
\def\Cpeff#1  {\ensuremath{\mathcal{C}_{#1}^{'\mathrm{(eff)}}}\xspace}       
\def\Ope#1    {\ensuremath{\mathcal{O}_{#1}}\xspace}                       
\def\Opep#1   {\ensuremath{\mathcal{O}_{#1}^{'}}\xspace}                    

\newcommand{\ket}[1]{\ensuremath{|#1\rangle}}              

\newcommand{\tev}{\ifthenelse{\boolean{inbibliography}}{\ensuremath{~T\kern -0.05em eV}\xspace}{\ensuremath{\mathrm{\,Te\kern -0.1em V}}}\xspace}
\newcommand{\gev}{\ensuremath{\mathrm{\,Ge\kern -0.1em V}}\xspace}
\newcommand{\mev}{\ensuremath{\mathrm{\,Me\kern -0.1em V}}\xspace}
\newcommand{\kev}{\ensuremath{\mathrm{\,ke\kern -0.1em V}}\xspace}
\newcommand{\ev}{\ensuremath{\mathrm{\,e\kern -0.1em V}}\xspace}
\newcommand{\gevc}{\ensuremath{{\mathrm{\,Ge\kern -0.1em V\!/}c}}\xspace}
\newcommand{\mevc}{\ensuremath{{\mathrm{\,Me\kern -0.1em V\!/}c}}\xspace}
\newcommand{\gevcc}{\ensuremath{{\mathrm{\,Ge\kern -0.1em V\!/}c^2}}\xspace}
\newcommand{\gevgevcccc}{\ensuremath{{\mathrm{\,Ge\kern -0.1em V^2\!/}c^4}}\xspace}
\newcommand{\mevcc}{\ensuremath{{\mathrm{\,Me\kern -0.1em V\!/}c^2}}\xspace}

\def\gsim{{~\raise.15em\hbox{$>$}\kern-.85em
          \lower.35em\hbox{$\sim$}~}\xspace}
\def\lsim{{~\raise.15em\hbox{$<$}\kern-.85em
          \lower.35em\hbox{$\sim$}~}\xspace}

\def\tell1  {TELL1\xspace}
\def\ukl1   {UKL1\xspace}

